\documentclass[useAMS,usenatbib]{mn2e}

\usepackage[version=3]{mhchem}
\usepackage{graphicx}

\begin{document}

   \title{The unstable \ce{CO2} feedback cycle on ocean planets}
   
   \author[D. Kitzmann et al.]{D. Kitzmann,$^1$\thanks{daniel.kitzmann@csh.unibe.ch}
	   Y. Alibert,$^1$\thanks{On leave from CNRS, Observatoire de Besan\c con, 41 avenue de l'Observatoire, 25000 Besan\c con, France}
           M. Godolt,$^2$
           J.~L. Grenfell,$^2$
           K. Heng,$^1$\newauthor
           A.~B.~C. Patzer,$^3$
           H. Rauer,$^{2,3}$
           B. Stracke,$^2$
           P. von Paris $^{4,5}$\\
           $^1$ Physikalisches Institut \& Center for Space and Habitability, Universit\"at Bern, 3012 Bern, Switzerland\\
           $^2$ Institut f\"ur Planetenforschung, Deutsches Zentrum f\"ur Luft- und Raumfahrt (DLR), 12489 Berlin, Germany\\
           $^3$ Zentrum f\"ur Astronomie und Astrophysik, Technische Universit\"at Berlin, 10623 Berlin, Germany\\
           $^4$ Univ. Bordeaux, LAB, UMR 5804, F-33270, Floirac, France\\
           $^5$ CNRS, LAB, UMR 5804, F-33270, Floirac, France           
         }

\maketitle

\begin{abstract}
Ocean planets are volatile rich planets, not present in our Solar System, which are thought to be dominated by deep, global oceans. This results in 
the formation of high-pressure water ice, separating the planetary crust from the liquid ocean and, thus, also from the atmosphere. Therefore, 
instead of a 
carbonate-silicate cycle like on the Earth, the atmospheric carbon dioxide concentration is governed by the capability of the ocean to dissolve 
carbon dioxide (\ce{CO2}).

In our study, we focus on the \ce{CO2} cycle between the atmosphere and the ocean which determines the atmospheric \ce{CO2} content. 
The atmospheric amount of \ce{CO2} is a fundamental quantity for assessing the potential habitability of the planet's surface because of its strong 
greenhouse effect, which determines the planetary surface temperature to a large degree. In contrast to the stabilising carbonate-silicate cycle 
regulating the long-term \ce{CO2} inventory of the Earth atmosphere, we find that the \ce{CO2} cycle feedback on ocean planets is negative and has 
strong 
destabilising effects on the planetary climate. By using a chemistry model for oceanic \ce{CO2} dissolution and an atmospheric model for exoplanets, 
we show that the \ce{CO2} feedback cycle can severely limit the extension of the habitable zone for ocean planets.
\end{abstract}

\nokeywords

\section{Introduction}

The question of habitability for exoplanets is usually linked to the concept of the so called habitable zone (HZ), i.e. the range of orbital 
distances around a host star, where a rocky/terrestrial planet can in principle maintain liquid water on its surface over an extended period of time. Liquid water is - to our 
current knowledge - one of the most important requirement for life as we know it \citep[e.g.][]{Lammer2009aA&ARv}. However, the classical HZ as 
introduced by 
\citet{Kasting1993} is strictly defined for an Earth-like planet, i.e. for a planet with a partially rocky surface, a water content of one Earth 
ocean, plate tectonics, and with all the geological cycles (in particular, the carbonate-silicate cycle) as on Earth. Given the diversity of the 
already known exoplanets, it is highly unlikely that the majority of the low-mass planets are entirely Earth-like and can be described within the 
concept of the classical HZ. 

One particular, highly interesting class of exoplanets are ocean planets (or waterworlds), i.e. planets with a much higher water content than 
Earth. Several candidates for massive water-rich planets have already been found. Due to their small orbital distances, none of these can 
be considered to be a potential habitable planet, though. The most well-studied example so far is the transiting super-Earth GJ1214 b 
\citep{Charbonneau2009Natur.462..891C}. Its mean density derived from observations suggests that roughly half the planet's bulk composition could be 
made of water. In general, the larger radii of ocean planets compared to more rocky planets of equal mass \citep{Leger:2004aa, Alibert:2014aa}, would 
make them very interesting targets for planet detection and characterization with, for example, \textit{CHEOPS} \citep[CHaracterising ExOPlanet 
Satellite, ][]{Fortier2014SPIE.9143E..2JF}, \textit{TESS} \citep[Transiting Exoplanet Survey Satellite, ][]{Ricker2014aSPIE.9143E..20R}, 
\textit{PLATO} \citep[Planetary Transits and Oscillations of stars, ][]{Rauer2014aExA....38..249R}, or the \textit{JWST} \citep[James Webb Space 
Telescope, ][]{Gardner2006aSSRv..123..485G}. Being completely covered by a deep water envelope, they fall within the apparent gap between 
terrestrial 
planets 
with a (partially) rocky surface and the gas giant planets with a high amount of hydrogen and helium.

In this study we focus on the potential habitability of low-mass deep ocean planets. First considered by \citet{Leger:2004aa}, these types of planets would 
form beyond the ice line within a protoplanetary disk, where they collect a large fraction of water as well as other volatiles as ices and later 
migrate inwards into the habitable zone \citep{Kuchner2003ApJ...596L.105K}. These considerations are supported by the results from planetary 
formation models, where migrated terrestrial low-mass planets with a large water inventory ($> 100$ Earth oceans) located in the classically defined 
habitable zone (HZ) are a very common result \citep{Alibert:2014aa, Alibert:2013aa, Thiabaud:2014aa}.

Studies on the internal structure of such planets by \citet{Leger:2004aa} and \citet{Alibert:2014aa} suggest that they will have ocean depths in 
excess of 100 km, which makes the presence of a continental landmass highly unlikely. Categorized as Class IV habitats by \citet{Lammer2009aA&ARv} 
these planets can offer one of the most important requirements for life: liquid water. Of course, such a large reservoir of water (\ce{H2O}) also 
influences climate processes. One important consequence is, for example, the formation of high-pressure water ice (ice phases VI and VII) at the 
bottom of the ocean, which prevents the immediate contact of the planetary crust with the liquid ocean. Thus, the well-known 
carbonate-silicate cycle \citep{Walker:1981aa, Kasting1993} is unable to operate as a long-term climate stabilisation mechanism regulating the 
atmospheric carbon dioxide content. \citet{Abbot2012ApJ...756..178A} also showed that below a landmass fraction of one percent, 
the usual seafloor weathering, which is an important part of the carbonate-silicate cycle, becomes inoperative, such that the stabilising 
weathering-climate feedback on ocean planets does not work even if the liquid ocean is still in contact with the planetary crust.

Since the ocean planet originates from beyond the ice line in the protoplanetary disk, the total amount of \ce{CO2} accumulated during 
its formation is also thought to be much higher than the Earth’s inventory (see e.g. \citet{Leger:2004aa}, \citet{Alibert:2014aa}, or 
\citet{Marboeuf2014A&A...570A..36M}). As 
discussed in \citet{Leger:2004aa}, the \ce{CO2} amount present above the high-pressure water ice layer may, however, be strongly limited by 
gravitational sequestration during the planetary formation (i.e. most of the \ce{CO2} is trapped below the water ice layer), clathrate formation, or 
atmospheric loss. In case of negligible atmospheric loss processes of \ce{CO2} or very inefficient formation of \ce{CO2}-\ce{H2O} clathrates, the total 
(initial) inventory of \ce{CO2} present in the ocean and the atmosphere remains constant over time, because \ce{CO2} cannot interact with the planetary 
interior via subduction or outgassing. Then, the partial pressure of \ce{CO2} in the atmosphere is purely determined by its partial dissolution in the 
liquid ocean.

Note that the occurrence of convection and plate tectonics within the ice mantle is quite uncertain. Studies by 
\citet{Fu2010ApJ...708.1326F} or \citet{Levi2014ApJ...792..125L} suggest the possibility of ice mantle convection on ocean planets under certain 
conditions. Convection depends on details like the temperature profile within the ice mantle, the planet's interior heat loss over time, or the 
viscosity of the high-pressure ice phases. Given the uncertainties in, for example, the experimental data of the ice viscosity or the detailed 
evolution of the planetary interior, it is currently not possible to constrain the occurrence of a convective ice mantle over the lifetime of the 
planet. Additionally, it was shown by \citet{Levi2014ApJ...792..125L} that methane (\ce{CH4}) could be transported through the convective ice mantle 
in the form of clathrates. To our knowledge, no theoretical or experimental data are available on clathrate formation for \ce{CO2} under the 
conditions encountered on an ocean planet. Assuming an efficient transport of carbon dioxide clathrates through the convective ice mantle would, in 
principle, connect the oceanic and atmospheric \ce{CO2} with the \ce{CO2} reservoir below the ice mantle. It is a priori not clear, if such a 
connection would work as a long-term stabilisation (as on Earth) or would even further enhance the \ce{CO2} effect discussed in this study. Due to 
these uncertainties, we assume that no transport of \ce{CO2} through the ice mantle occurs on the ocean planets in our study.

In this study we investigate the interaction of the ocean with the atmospheric \ce{CO2}. We present a potentially unstable \ce{CO2} cycle which, 
unlike the carbonate-silicate-cycle on Earth, acts as a destabilising climate feedback and might limit the overall habitability of an ocean planet. 
Oceanic dissolution of \ce{CO2} is also a well known effect in Earth climatology. The oceanic water of Earth has an overall \ce{CO2} content of about 
two bar \citep{Pierrehumbert2010ppc..book.....P}, which is much more than its actual corresponding atmospheric partial pressure of roughly $10^{-4}$ 
bar. The increased anthropogenic release of \ce{CO2} into the atmosphere results in an ocean acidification which will considerably lower the pH value 
of the Earth ocean within the next centuries \citep{Caldeira2003Natur.425..365C}. While this will have a strong effect on the maritime biosphere in 
the near future, the overall impact on the atmosphere and surface temperature is limited to a few degree kelvin because of the small size of the 
Earth oceans (compared to ocean planets) and the long-term stabilisation via the carbonate-silicate cycle. 

While ocean planets can in principle be more massive, we limit our present study to planets with the mass and radius of the Earth, orbiting a Sun-like star, as 
first example. Otherwise, we would need additional assumptions on the bulk composition and internal structure of the ocean planet to derive a 
corresponding mass-radius relationship. This, however, is not the topic of this study which focusses on the impact of the temperature-dependent dissolution of 
\ce{CO2} in the ocean on the planetary habitability. The results of this study are also directly applicable to ocean planets with higher masses and 
radii, though. We assume that any initially accumulated hydrogen and helium is already lost, which is a reasonable assumption for low-mass planets 
\citep{Alibert:2014aa, Lammer2008SSRv..139..399L}.

The details of the model for the oceanic \ce{CO2} dissolution are described in Sect. \ref{sect:co2_dissolution}. In Sect. \ref{sect:atmosphere_model} we discuss the 
atmospheric model and employ it to study the habitability of ocean planets

\section{Oceanic \ce{CO2} dissolution}
\label{sect:co2_dissolution}

Both the amount of liquid ocean water (i.e. the size of the 
global ocean) as well as the dissolved \ce{CO2} within the ocean (and thus also the atmospheric \ce{CO2} content) are in general a function of 
temperature. Thus, we derive the size of the liquid ocean and describe the chemistry for the dissolution of \ce{CO2} in ocean water in this section. We 
thereby do not account for loss processes of atmospheric \ce{CO2} to space or for the potential formation of \ce{CO2}-\ce{H2O} clathrates as both 
of them are hard to quantify without more additional assumptions.

\subsection{Ocean mass}

In the following, we assume that the ocean is isothermal with a temperature equal to the temperature at the ocean's surface to study the lower limit of the \ce{CO2} feedback impact on the atmospheres of low-mass ocean planets. This assumption results in the smallest possible ocean size. If, for example, an adiabatic temperature profile had instead been adopted \citep{Leger:2004aa}, the ocean would have been larger and able to store a larger quantity of \ce{CO2}. This on the other 
hand would amplify the proposed \ce{CO2} feedback mechanism as discussed below.

\begin{table}
\caption{Simon-Glatzel equation parameters for the melting curves of water ices.}
\label{symbols}
\begin{tabular}{@{}lcccccc}
\hline
 Ice & \multicolumn{2}{c}{Triple point parameters} & & \multicolumn{2}{c}{Power-law parameters}\\\cline{2-3}\cline{5-6}
 phase & $p_0$ (MPa) & $T_0$ (K)& & $a$ (MPa)& $c$\\
\hline
Ice VI$^1$ & 618.4 & 272.73 & & 661.4 & 4.69\\
Ice VII$^2$ & 2216 & 355 & & 534 & 5.22\\
\hline
\end{tabular}
\medskip
$^1$ \citet{Choukroun2007JChPh.127l4506C}
$^2$ \citet{Alibert:2014aa}
\label{table:melting_curves}
\end{table}

Assuming an isothermal ocean, its maximal mass is constrained by the maximum pressure of liquid water, given by the temperature and 
pressure-dependent melting curves of water ice phases VI and VII. This maximum pressure $p$ is described as a function of temperature $T$ by 
the empirically derived Simon-Glatzel equation \citep{simon1929bemerkungen}
\begin{equation}
  p = p_0 + a \left( \left(\frac{T}{T_0}\right)^{c} - 1 \right)
\end{equation}
where $p_0$ and $T_0$ are reference values (usually the corresponding triple points) and $a$ and $c$ are power-law parameters obtained from 
experimental data. The corresponding parameters for the equation are summarised in Table \ref{table:melting_curves} and have been adopted from 
\citet{Alibert:2014aa} for ice phase VII and from \citet{Choukroun2007JChPh.127l4506C} for ice VI.

\begin{figure}
\resizebox{\hsize}{!}{\includegraphics{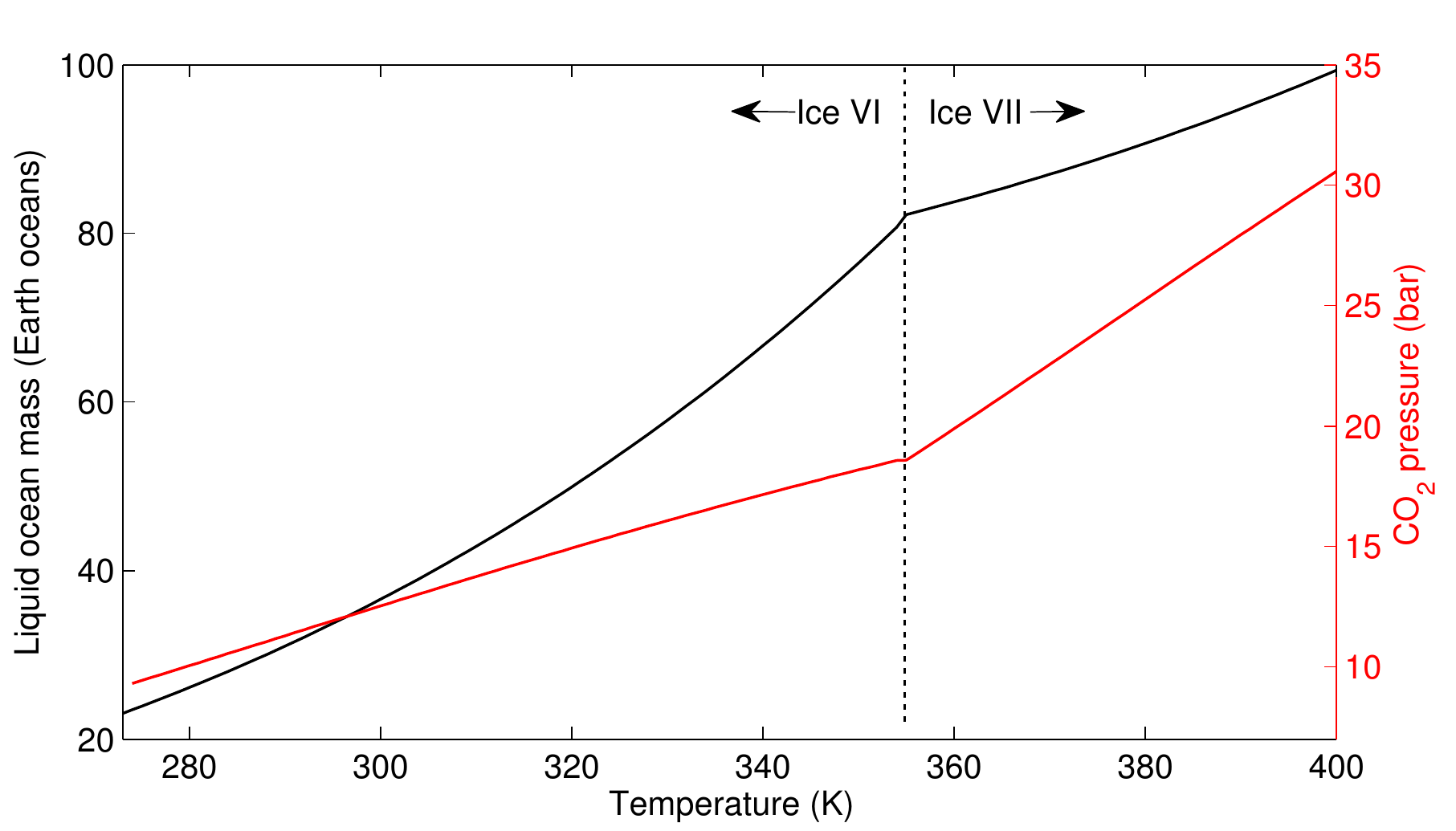}}
\caption{Mass of the liquid ocean and total atmospheric \ce{CO2} pressure as a function of isothermal ocean temperature. The atmospheric \ce{CO2} 
partial
pressure (red line) is shown for a total \ce{CO2} inventory of $10^{21}$ kg and $10^{18}$ kg of \ce{N2}. The vertical dotted line indicates the 
boundary between the formation of the ice phases VI and VII at the ocean’s bottom.}
\label{fig:ocean_mass}
\end{figure}

The ocean mass is shown in Fig. \ref{fig:ocean_mass} as a function of (isothermal) 
temperature. It varies considerably from about 25 to 100 Earth oceans between 273 K and 400 K. Thus, at low temperatures, less liquid water is 
available to dissolve the atmospheric \ce{CO2}.

\subsection{\ce{CO2} dissolution model}
\label{subsec:dissolution_model}

Following the description of \citet{Pierrehumbert2010ppc..book.....P}, we derive the atmospheric \ce{CO2} content by taking into account the 
dissolution of \ce{CO2} within the ocean. In equilibrium, the atmospheric partial pressure of \ce{CO2(g)}, $p_\ce{CO2(g)}$, and the concentration of 
\ce{CO2(aq)} dissolved in the ocean ($c_\ce{CO2(aq)}$) are linked by Henry's law 
\begin{equation}
  p_\ce{CO2(g)} = K_H(T) \cdot c_\ce{CO2(aq)}
\end{equation}
with the temperature-dependent Henry's law constant $K_H(T)$. Additional reactions occurring within the ocean convert the dissolved \ce{CO2(aq)} into 
hydrogen carbonate (\ce{HCO_3^-}) and carbonate (\ce{CO_3^{2-}})
\begin{equation}
  \ce{CO2(aq) + H2O(l) <=> HCO3^-(aq) + H^+(aq)}  \ ,
  \label{eq:reaction_1}
\end{equation}
\begin{equation} 
  \ce{HCO3^-(aq) <=> CO3^{2-}(aq) + H^+(aq)}  \ .
  \label{eq:reaction_2}
\end{equation}
The hydrogen ion \ce{H^+} is additionally linked to the hydroxide ion \ce{OH^-} via the dissociation reaction of water
\begin{equation}
  \ce{H2O(l) <=> H^+(aq) + OH^-(aq)} \ .
\end{equation}

Assuming chemical equilibrium, we solve this system under the constraint of charge balance 
\begin{equation}
  \left[\ce{H^+}\right] = \left[\ce{HCO_3^-}\right] + 2 \left[\ce{CO3^{2-}}\right] + \left[\ce{OH^-}\right]
  \label{eq:charge_balance}
\end{equation}
by using a Newton-Raphson method, iterating the mole concentration $\left[\ce{H+}\right]$. 
Rate constants for the chemical reactions and Henry's law have been taken from \citet{Pierrehumbert2010ppc..book.....P}. The temperature dependence 
of the reaction rates are described via the Arrhenius equation. The rate coefficients used here are defined for an Earth-like 
oceanic salinity of 30\%. The salinity can have a large effect on the rates of the reactions \ref{eq:reaction_1} and \ref{eq:reaction_2}. Tests with 
low-salinity rate constants (taken from the supplementary material of \citet{Pierrehumbert2010ppc..book.....P}) indicate, that our results are 
mostly insensitive to the oceanic salinity. In the cases studied in this paper, most of the carbon is stored as dissolved \ce{CO2} in the ocean, such 
that the reactions \ref{eq:reaction_1} and \ref{eq:reaction_2} are only of minor importance.

Note that 
$\left[\ce{H+}\right]$, when expressed in units of moles per litre, is associated with the pH value of the ocean via
\begin{equation}
  \mathrm{pH}_\mathrm{ocean} = -\log_{10} \left[\ce{H+}\right] \ .
\end{equation}

Given a total \ce{CO2} mass inventory $m_{\ce{CO2_{,tot}}}$, a second Newton-Raphson method is applied to find the atmospheric mass of \ce{CO2} 
($m_{\ce{CO2(g)}}$) for which the conservation of mass 
\begin{equation}
  m_{\ce{CO2_{,tot}}} = m_{\ce{CO2(g)}} + m_{\ce{CO2(aq)}}
\end{equation}
is achieved, where $m_{\ce{CO2(aq)}}$ is determined by the chemical system described above as a function of $p_\ce{CO2(g)}$. 
Atmospheric mass and partial pressure of \ce{CO2(g)} are related via the condition of hydrostatic equilibrium.

We do not consider the presence of positive calcium ions (\ce{Ca^{2+}}) in the charge balance. On planets like Earth, these ions can originate from 
the dissolution of limestone (\ce{CaCO3}) in water
\begin{equation}
  \ce{CaCO3(s) <=> Ca^{2+}(aq) + CO3^{2-}(aq)} \ .
\end{equation}
Since, however, the liquid water on the planets under consideration is not in contact with a rocky surface, this source for the \ce{Ca^{2+}} ions 
does not exist. Note that adding \ce{Ca^{2+}} ions via dissolved limestone would strongly increase the capability of the ocean to 
take up \ce{CO2} and, thereby, enhance the \ce{CO2} feedback cycle\footnote{This is of course also true for all other influences affecting the 
charge balance in Eq. \ref{eq:charge_balance}.}.

\subsection{Results}
\label{subsec:co2_dissolution_results}

\begin{figure}
\includegraphics[width=84mm]{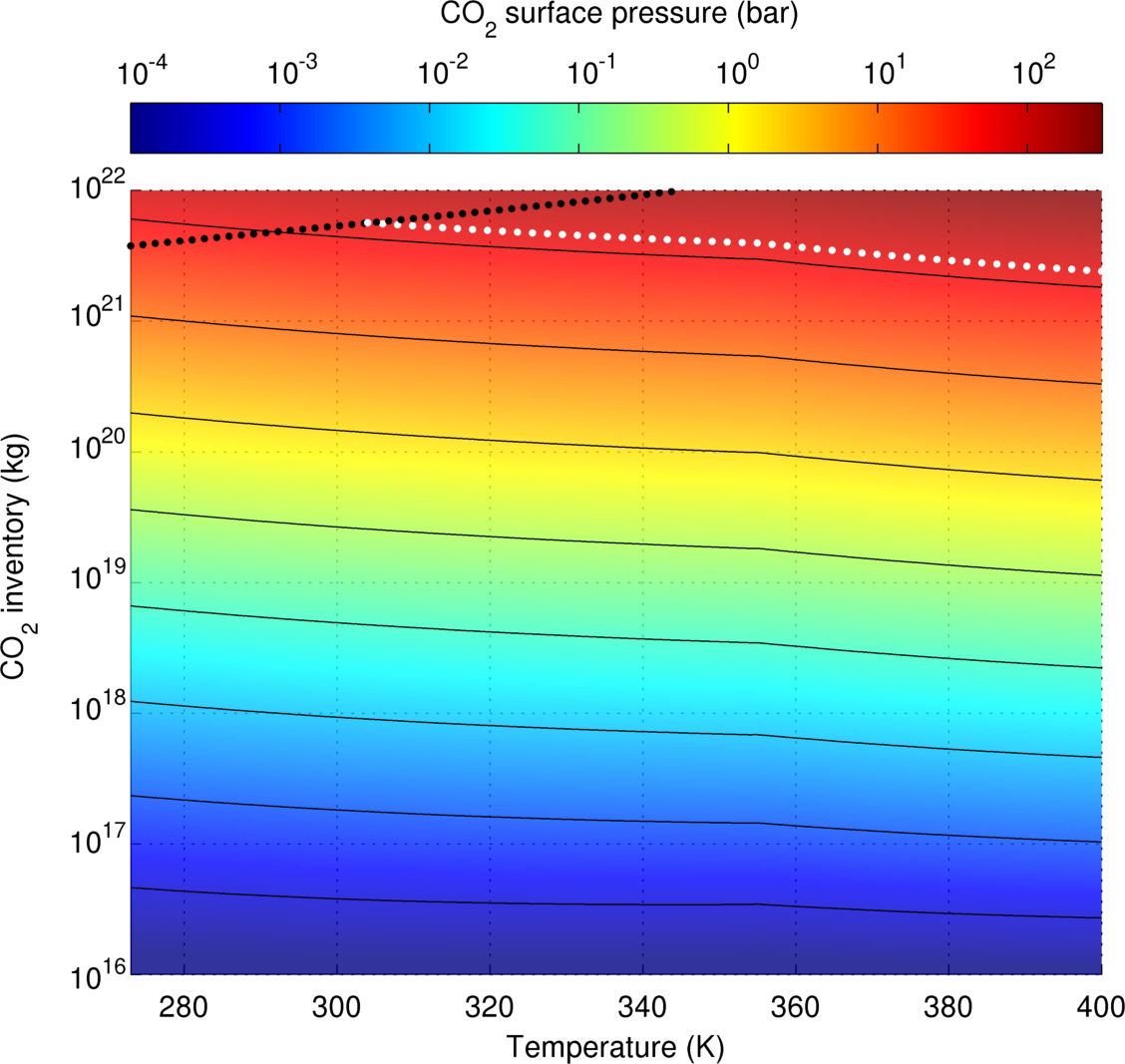}
\caption{Atmospheric partial pressure of \ce{CO2} as a function of ocean temperature and total \ce{CO2} inventory. The amount of \ce{N2} in the 
atmosphere is fixed at $10^{18}$ kg in each case. The solid, black lines represent contour lines for constant \ce{CO2} pressures. The condensation 
limit for 
\ce{CO2} is indicated by the dotted line. Atmospheric partial pressures of \ce{CO2} above that line are thermodynamically unstable and would 
condense out. The white, dotted line marks the critical point limit of \ce{CO2}. Above this limit, the treatment of the equilibrium chemistry 
becomes unsuitable. }
\label{fig:co2_pressure}
\end{figure}

Figure \ref{fig:ocean_mass} shows the resulting atmospheric partial pressure \ce{CO2(g)} as a function of isothermal ocean temperature for a total 
inventory of $10^{21}$ kg as a particular example for an intermidate value of the range of \ce{CO2} inventories considered in the following. The 
atmospheric mass of nitrogen (\ce{N2}) is fixed at an approximately Earth-like value of $10^{18}$ kg.
For this example, the surface pressure of \ce{CO2(g)} increases from about 9 bar at 273 K to more than 30 bar at 400 K. Taking into account 
the results of planetary population synthesis models from e.g. \citet{Marboeuf2014A&A...570A..36M} to constrain the overall amount of \ce{CO2} 
accumulated from the protoplanetary disc 
for low-mass planets which have migrated into the habitable zone, we vary the total \ce{CO2} inventory between $10^{16}$~kg and $10^{24}$~kg. The 
results are presented in Fig. \ref{fig:co2_pressure}, where a \ce{N2} inventory of $10^{18}$~kg is used again in each case. 

We note that the results in Fig. \ref{fig:co2_pressure} for very high total \ce{CO2} inventories are physically unrealistic as they predict 
atmospheric partial pressures of more than 100 bar. Such a large amount of \ce{CO2(g)} would partially condense onto the surface as liquid \ce{CO2}. 
According to the vapour pressure curve \citep{AmbroseTF9565200772}, the highest partial pressure of \ce{CO2(g)} at 400 K is roughly 600 bar while at 
273 K only about 34.7 bar can be stable in the atmosphere in gaseous form. The condensation limit is indicated by the dotted, black line in Fig. 
\ref{fig:co2_pressure}. Atmospheric partial pressures of \ce{CO2} above that line are thermodynamically unstable and would condense out, removing 
it from the atmosphere \citep[see also][]{vonParis2013A&A...549A..94V}. 
Additionally, some of the results shown in Fig \ref{fig:co2_pressure} lie 
above the critical point of \ce{CO2}, which is located at a temperature of 304.25 K and pressure of 73.8 bar. Above the critical point, the 
description of the equilibrium 
chemistry from Sect. \ref{subsec:dissolution_model} is probably unsuitable and the corresponding results should be treated with great care. The 
critical point limit is indicated by the dotted, white line in Fig. \ref{fig:co2_pressure}. We note that none of these unrealistic values 
are 
used for the atmospheric model calculations in Sect. \ref{sect:atmosphere_model}.

In general, the \ce{CO2(g)} partial pressure varies over several orders of magnitudes across the parameter space. Even though, the ocean is 
smaller at low temperatures and, thus, less liquid water is available to dissolve the atmospheric \ce{CO2}, the temperature dependence of the 
dissolution is stronger. The ability of the ocean to dissolve \ce{CO2} decreases with increasing temperature, which leads to a positive, and thus 
potentially destabilizing, feedback cycle within the atmosphere. The partial pressure of \ce{CO2(g)} likewise decreases with temperature owing to the 
ability of the ocean to dissolve a larger amount of \ce{CO2} at lower temperatures. This, again results in a positive feedback cycle. 

Thus, at low 
surface temperatures, where \ce{CO2} would be required as an atmospheric greenhouse gas to heat the planetary surface, the ocean 
binds an increasing amount of \ce{CO2} and, thereby, removes the important greenhouse gas from the atmosphere. For higher temperatures on the other 
hand, less \ce{CO2} can be dissolved in the ocean and starts to accumulate in the atmosphere, contributing to the greenhouse effect and further 
increasing the surface temperature. In contrast to the stabilising carbonate-silicate cycle operating on Earth, this therefore induces positive 
feedback cycles, which could lead to potentially unstable situations: an atmospheric freeze-out with a runaway glaciation or a runaway greenhouse 
effect, for example. This is directly comparable to the water vapour feedback where the atmospheric content of \ce{H2O} increases strongly with 
surface temperature, which may finally result in a runaway greenhouse process \citep{Komabayashi1967, Kasting1988Icar}. These positive \ce{CO2} 
feedback cycles will obviously have important implications for the potential habitability of an ocean planet. This effect has also been noted in the 
appendix of \citet{Wordsworth2013ApJ...778..154W} as a possibly important mechanism on volatile-rich planets.

The highest surface temperatures considered in this study are limited to 400 K, assuming that this is the limit for microorganisms to survive and thus 
a potential limit for the presence of life as we know it \citep{Cockell1999P&SS...47.1487C, Lammer2009aA&ARv, 
Rothschild2001Natur.409.1092R}. The lowest surface temperature for the limit of potential habitability is considered to be the freezing point of water 
(273 K). Note that 3D studies show that the global mean surface temperature of 273 K may still allow for liquid water in the equatorial region 
\citep[see e.g.][]{Wolf2013AsBio..13..656W, Kunze2014P&SS...98...77K}

\section{Atmospheric model calculations}
\label{sect:atmosphere_model}

Since the planetary surface temperature depends - amongst other things - on the atmospheric amount of \ce{CO2} which, as described above, is a 
function of the ocean’s temperature, we use an atmospheric model coupled with the \ce{CO2(aq)} dissolution model to constrain the potential 
habitability of the ocean planet. The one-dimensional, radiative-convective atmospheric model was developed for terrestrial planets and is described 
in the next subsection. The results from the model calculations using the dissolution model and the atmospheric model are shown in Sect. 
\ref{subsec:model_results}. Focussing on the \ce{CO2} feedback mechanism, we present the model calculations for a G2V host star: the Sun. However, this study can of course be applied to other central stars straightforwardly. 
The atmospheric composition for this study is restricted to a mixture of water, carbon dioxide, and molecular nitrogen. The atmospheric 
amount of \ce{N2} remains fixed at an approximately Earth-like value of 10$^{18}$ kg. Nitrogen, unless present in very high amounts, has a very 
limited effect on the resulting surface temperature \citep{vonparis2013P&SS...82..149V, Wordsworth2013ApJ...778..154W}.

\subsection{Atmospheric model description}

For the atmospheric model calculations we use a modified version of the one-dimensional radiative-convective atmospheric model extensively described 
in \citet{Kitzmann2010A&A...511A..66K} and \citet{vonParis2010A&A...522A..23V}. The model has been updated in \citet{StrackePSS2015submitted} to 
include new absorption coefficients in the shortwave part of the radiative transfer based on the HITEMP2010 database 
\citep{Rothman2010JQSRT.111.2139R}. The model considers \ce{CO2}, \ce{H2O}, and \ce{N2} as 
atmospheric gases, neglecting other, potentially important, radiative trace species which might be present, such as \ce{O2}, \ce{SO2}, \ce{O3}, or 
\ce{CH4}. 

Nitrogen and \ce{CO2} are assumed to be well mixed throughout the atmosphere. Water vapour profiles for low surface temperatures are calculated by 
the fixed relative humidity profile of \citet{Manabe67} through the troposphere. For the water-rich atmospheres at high stellar insolations, a 
relative humidity of unity is used \citep{Kasting1993, Kopparapu2013ApJ...765..131K}. Above the cold trap, the water profile is set to an isoprofile 
using the \ce{H2O} concentration at the cold trap. 

The temperature profile is calculated from the requirement of radiative equilibrium by a time-stepping approach, as well as performing a convective 
adjustment \citep{Manabe1964JAtS...21..361M}, if necessary. The convective lapse rate is assumed to be adiabatic, taking into account the 
condensation of \ce{H2O} and \ce{CO2} \citep{Kasting1993,vonParis2010A&A...522A..23V}.

As described in \citet{vonParis2010A&A...522A..23V}, the cloud-free radiative transfer of the atmospheric model is split into two wavelength regimes: 
a shortwave part, dealing with the wavelength range of the incident stellar radiation and the longwave part for the treatment of the atmospheric 
thermal radiation. Both parts use the correlated-$k$ method to describe the gaseous opacity. For the shortwave radiative transfer, 38 spectral 
intervals between 0.238 $\mu$m and 4.55 $\mu$m are used, in the longwave part 25 bands from 1 $\mu$m to 500 $\mu$m 
\citep{vonParis2010A&A...522A..23V}. In this wavelength region, only absorption by \ce{CO2} and \ce{H2O} is taken into account. Rayleigh scattering 
is considered for \ce{CO2}, \ce{N2}, and \ce{H2O} \citep{vonParis2010A&A...522A..23V}. Continuum absorption is taken into account for \ce{CO2} in the 
longwave \citep{vonParis2010A&A...522A..23V} and \ce{H2O} in both wavelength regions \citep{Clough1989AtmRe..23..229C}. The equation of radiative 
transfer is solved by two-stream methods \citep{Toon1989JGR....9416287T}. In the shortwave region, a $\delta$-Eddington two-stream method is employed, 
while in the longwave part a hemispheric-mean two-stream radiative transfer is used \citep{Kitzmann2010A&A...511A..66K}. The spectrum of the Sun 
describing the incident stellar radiation has been compiled from data published by \citet{Gueymard2004} (see \citet{Kitzmann2010A&A...511A..66K} 
for details). A planetary surface albedo of 0.06 for liquid water is used throughout this study \citep{Scinocca2006JCli...19.6314L}.

\subsection{Results} 
\label{subsec:model_results} 

In principle, it is possible to couple the chemical \ce{CO2} dissolution model from Sect. \ref{subsec:dissolution_model} directly with the atmospheric 
model to calculate the atmospheric content of \ce{CO2} interactively every time step as a function of surface temperature. However, this approach does 
not reliably converge in every case since one of the main atmospheric greenhouse gases is changed constantly within a positive feedback cycle. This 
easily leads to runaway scenarios, 
where \ce{CO2} either accumulates in the atmosphere or is removed until the atmosphere freezes out. Instead of a direct coupling, we use a slightly 
different approach in this study. 

Since the desired surface temperatures are known a priori (273 K and 400 K), the atmospheric content of \ce{CO2} is 
fixed at the corresponding value from the \ce{CO2} dissolution model shown in Fig. \ref{fig:co2_pressure}. With the fixed \ce{CO2} atmospheric 
partial pressure, the atmospheric model is used to find the corresponding orbital distance where the atmosphere is in thermal equilibrium for the 
chosen surface temperatures and \ce{CO2} inventories.

\begin{figure}
\resizebox{\hsize}{!}{\includegraphics{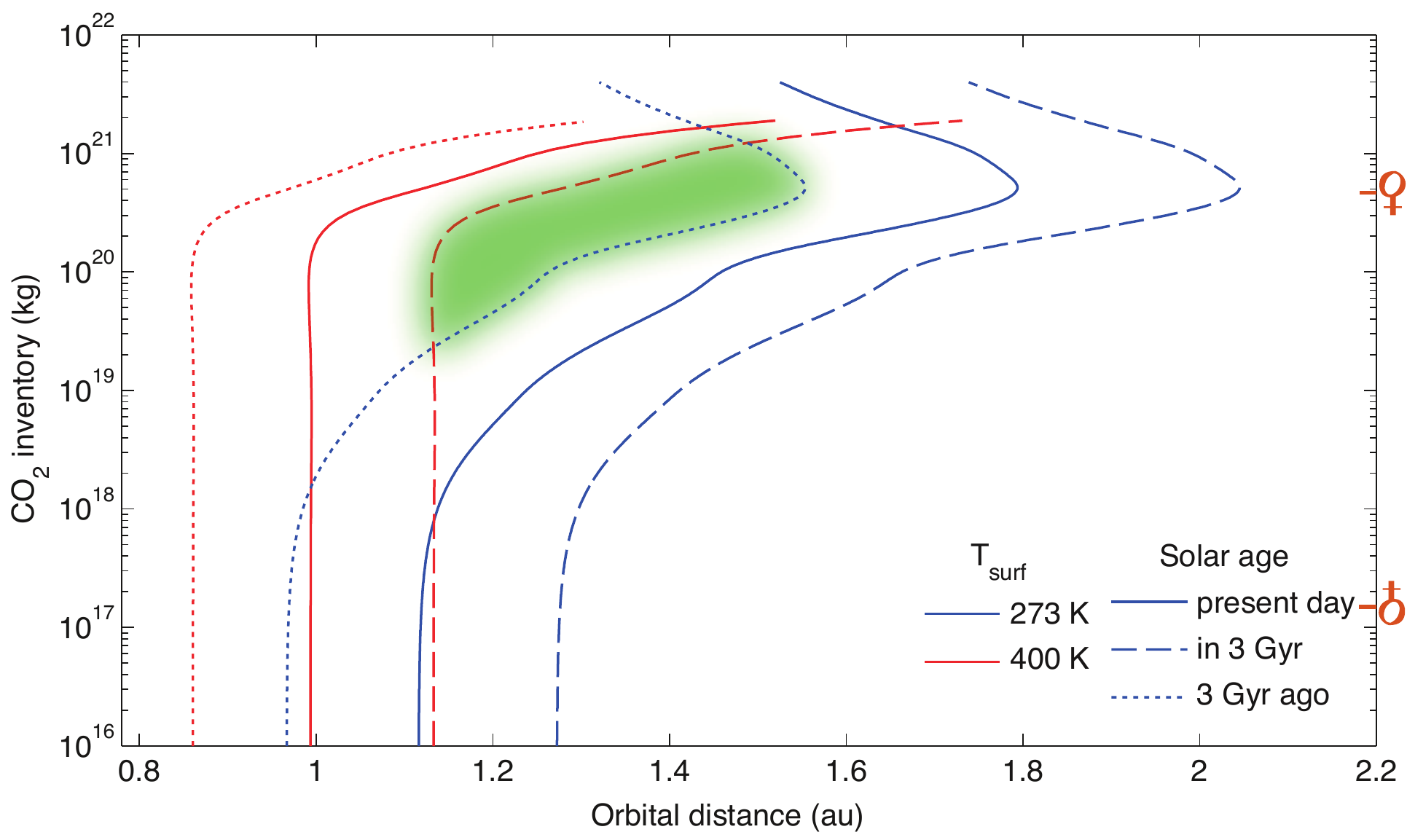}}
\caption{Orbital distances for habitable surface conditions. Coloured lines represent surface temperatures T$_\mathrm{surf}$ of 273 K (blue) and 400 
K (red). Line styles indicate different stellar ages of the Sun. The green area shows the distances and \ce{CO2} inventories for which the planet 
remains habitable over the entire time period (6 Gyr). The corresponding inventories of \ce{CO2(g)} for Venus and \ce{CO2(g)}+\ce{CO2(aq)} for Earth 
are indicated at the right-hand axis for comparison.}
\label{fig:distance}
\end{figure}

The resulting orbital distances for surface temperatures of 273 K and 400 K are shown in Fig. \ref{fig:distance}. We 
additionally compute the distances for different stellar ages, taking into account the change of the solar luminosity over time 
\citep{Gough1981SoPh...74...21G}. The results 
depicted in Fig. \ref{fig:distance} suggest that the range of distances for which the ocean planet remains habitable over an extended period of time 
(6 Gyr) is quite small. It is strongly restricted to \ce{CO2} inventories between 10$^{19}$ kg and 10$^{21}$ kg and to 
orbital distances between 1.1 au and 1.5 au. Thus, without the stabilising carbonate silicate cycle, the habitable zone of ocean planets might be 
much smaller than predicted by the concept of the classical HZ \citep{Kasting1993, Kopparapu2013ApJ...765..131K}.

At large orbital distances the results are similar to the well-known maximum \ce{CO2} greenhouse effect \citep{Kasting1993, 
Kopparapu2013ApJ...765..131K}. In this case, a maximum partial pressure of \ce{CO2} exists where the greenhouse effect still outweighs the Rayleigh 
scattering, yielding a net heating. For larger partial pressures, however, Rayleigh scattering dominates. Thus, the planet must be located 
closer to the host star to keep the surface temperature above the freezing point of water. At a temperature of 273 K, the total 
amount of \ce{CO2(g)} which can be present in the atmosphere is restricted to about 34.7 bar (equivalent to roughly $3.7\cdot10^{21}$ kg of total 
\ce{CO2}), because any additional \ce{CO2} in excess of that partial pressure would condense onto the surface, forming a layer of liquid \ce{CO2} 
floating on top of the ocean. 

For small \ce{CO2} inventories, the planetary conditions at inner distances are mostly determined by the usual water 
vapour feedback cycle which might eventually lead to a runaway greenhouse effect \citep[however, see][for new results 
concerning the runaway greenhouse effect]{StrackePSS2015submitted}. In contrast to that, the atmospheric partial pressure of \ce{CO2} 
for larger inventories is sufficient to increase the surface temperature above 400 K already far beyond the current Earth orbit. The positive 
feedback cycle results here in an atmospheric \ce{CO2} content larger than 30 bar, rendering the planetary surface uninhabitable.   
These findings imply that ocean planets are most likely continuously habitable only for very specific conditions. The positive feedback cycle of the 
oceanic \ce{CO2} dissolution severely constrains the potential habitability of these types of planets.

\begin{figure}
\resizebox{\hsize}{!}{\includegraphics{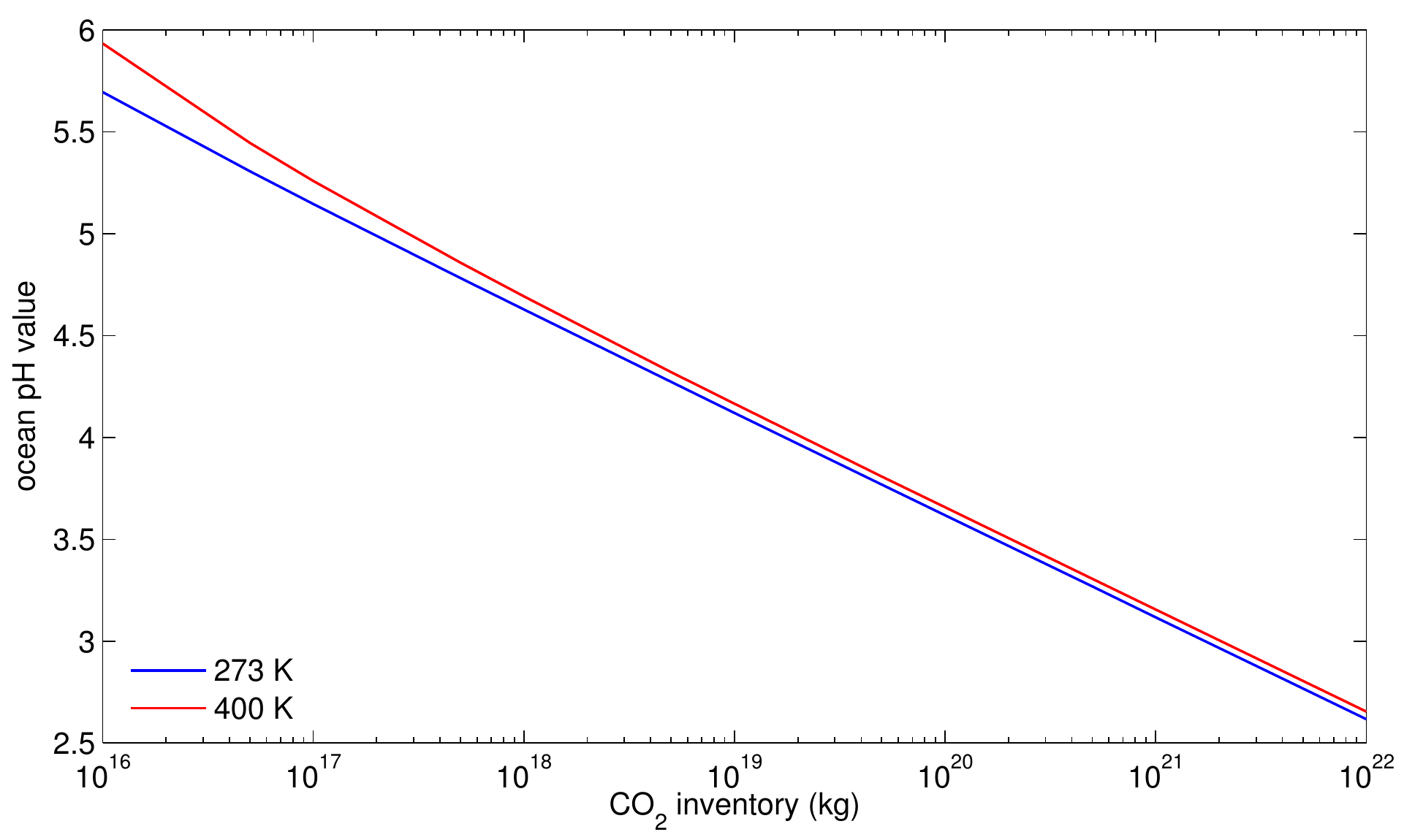}}
\caption{Oceanic pH values as a function of total \ce{CO2} inventory. The red line denotes surface temperatures of 400 K, the blue one 273 K.}
\label{fig:ph_value}
\end{figure}

The dissolution of \ce{CO2} in the ocean also has a strong impact on its pH value. Figure \ref{fig:ph_value} shows the resulting pH values for the 
two different surface temperatures as a function of total \ce{CO2} inventory. With increasing \ce{CO2} content, more \ce{CO2(aq)} is dissolved in the 
ocean which considerably lowers the pH value by the release of \ce{H^+} ions needed to meet the requirement of charge balance (Eq. 
\eqref{eq:charge_balance}). Starting at pH values of about six, it decreases to values of roughly 2.5 for the considered \ce{CO2} inventories. This 
may, additionally to the temperature effect, have an impact on the potential habitability of the liquid ocean 
\citep{Rothschild2001Natur.409.1092R, Lammer2009aA&ARv}.
The pH values for surface temperatures of 400 K are overall lower compared to the one for 273 K. This is related to the fact that at 273 K - even 
though the liquid ocean is smaller - more \ce{CO2} is dissolved in the ocean. Note, that even at very low \ce{CO2} inventories, the ocean of the 
waterworld has already a much lower pH value than the Earth ocean, with a pH value of about eight. This is caused by the absence of limestone in 
contact with the ocean planet's liquid water which, when dissolved, would reduce the acidity of the ocean, buffering the pH 
\citep{Pierrehumbert2010ppc..book.....P}.

\section{Summary}

In this study we investigated the potential habitability of extrasolar ocean planets. We focused on the \ce{CO2} cycle where the atmospheric \ce{CO2} 
can be partially dissolved in the deep water ocean. The high pressure at the ocean's bottom results in the formation of high-pressure water ices which 
separate the liquid water from the planetary crust, thus rendering the usual, stabilising carbonate-silicate cycle inoperable.

We used a chemical equilibrium chemistry to calculate the temperature-dependent dissolution of atmospheric \ce{CO2} in the liquid ocean, taking also 
the change of the ocean's size with temperature into account. The results from the dissolution model imply that the \ce{CO2} cycle operating on ocean 
planets is positive and, thus, potentially destabilising. The ability of the ocean to dissolve \ce{CO2} increases with decreasing temperature. Thus, 
at low surface temperatures the ocean can bind an increasing amount of \ce{CO2} which removes the important 
greenhouse gas from the atmosphere where it is needed to retain habitable conditions on the surface. For higher temperatures on the other 
hand, less \ce{CO2} can be dissolved in the ocean and starts to accumulate in the atmosphere which results in the build-up of a massive \ce{CO2} 
dominated atmosphere. This greatly contributes to the greenhouse effect, further increasing the surface temperature.

The results of the dissolution model were then used in an atmospheric model for exoplanets to calculate the distances of the ocean planets to their host 
star where the surface still remains habitable for an extended period of time. We showed that the range of distances is very restricted in terms of 
the total \ce{CO2} inventory of an ocean planet.

These findings imply that ocean planets are most likely continuously habitable only for very specific conditions. The positive feedback cycle of the 
oceanic \ce{CO2} dissolution severely constrains the potential habitability of these types of planets. While this study is focused on
carbon dioxide, the same temperature dependent solubility and therefore positive feedback cycles can also be obtained for other important greenhouse 
gases, such as methane, for example. Thus, our results have important implications for the general question of habitability for exoplanets because the 
classical concept of the habitable zone for Earth-like planets which implicitly assumes an operating carbonate-silicate cycle is obviously not 
applicable to water-rich exoplanets.
We note, however, that other - potentially important factors - might influence the results presented in this study. This includes e.g. the potential 
formation of \ce{H2O-CO2} clathrates or the transport of \ce{CO2} through the ice layer via clathrate diffusion or convection processes within the 
ice mantle all of which can  affect the \ce{CO2} content in the ocean and the atmosphere.

Finally, we also want to point out another possible constraint for the habitability of ocean planets. Since these planets migrate from beyond the 
protoplanetary disk’s ice line, the water can be initially in a frozen state \citep{Leger:2004aa} and has to be melted before any secondary 
atmosphere can be formed. 
Water ice, however, has a high albedo and, thus, reflects a large fraction of the incident radiation from a solar-type star. In the case that the 
primordial hydrogen-helium atmosphere \citep{Pierrehumbert2011ApJ}, or other processes, such as impacts, the energy acquired during the accretion 
phase within the disk, or a \ce{CO2}/\ce{CH4} atmosphere outgassed through the ice mantle \citep{Levi2014ApJ...792..125L},
did not result in an initial melting of the water ice, the stellar insolation must then be very high to melt the icy surface and form a liquid ocean. 
Assuming a conservative ice albedo of 0.5 and present day solar luminosity, the planet must be located at about 0.74 au to attain a global mean 
surface temperature of 273 K. At this distance, however, the planet, once the liquid ocean starts to form, would presently undergo a 
\ce{H2O} runaway 
greenhouse scenario \citep{Komabayashi1967, Kasting1988Icar}, rendering it uninhabitable. Being a potential strong constraint on the 
habitability of ocean planets, this process warrants 
further investigation, especially with regard to the ice albedo feedback for different stellar types \citep{vonparis2013AsBio..13..899V}.

\section*{Acknowledgements}
This work has been carried out within the frame of the National Centre for Competence in Research PlanetS supported by the Swiss National Science 
Foundation. The authors acknowledge the financial support of the SNSF.

This study has also received financial support from the French State in the frame of the ``Investments for the future'' Programme IdEx Bordeaux, 
reference ANR-10-IDEX-03-02.

M. Godolt acknowledges funding by the Helmholtz Foundation via the Helmholtz Postdoc Project ``Atmospheric dynamics and photochemistry of Super-Earth 
planets''.

The authors are grateful to the referee R. Pierrehumbert for his suggestions and advices.

\bibliographystyle{aa} 
\bibliography{references}

\end{document}